\documentclass[a4paper,12pt]{article}
\usepackage{graphicx,amssymb,bm}

\newcommand{\bea}{\begin{eqnarray}}
\newcommand{\ena}{\end{eqnarray}}
\newcommand{\vs}[1]{\vspace{#1 mm}}
\newcommand{\hs}[1]{\hspace{#1 mm}}
\renewcommand{\a}{\alpha}

\newcommand{\nn}{\nonumber\\}
\newcommand{\p}[1]{(\ref{#1})}

\begin{document}
\topmargin -10mm
\oddsidemargin 0mm

\renewcommand{\thefootnote}{\fnsymbol{footnote}}

\begin{titlepage}
\setcounter{page}{0}
\begin{flushright}
OU-HET 422 \\
USTC-ICTS 1 \\
hep-th/0210206
\end{flushright}

\vs{5}
\begin{center}
{\Large \bf NCOS and D-branes in Time-dependent Backgrounds}
%{\Large\bf S-Duality and Time-Dependent Noncommutative Theories}
\vs{10}

{\large Rong-Gen Cai\footnote{e-mail address:
cairg@itp.ac.cn}$^1$, Jian-Xin Lu\footnote{e-mail
address:jxlu@ustc.edu.cn}$^2$ and
Nobuyoshi Ohta\footnote{e-mail address: ohta@phys.sci.osaka-u.ac.jp}$^3$} \\
\vs{5}
{\em
$^1$ Institute of Theoretical Physics, Chinese Academy of Sciences,
P. O. Box 2735, Beijing 100080, China \\
$^2$ Interdisciplinary Center for Theoretical Study\\
 University of
Science and Technology of China, Hefei, Anhui 230026, China \\
$^2$ Michigan Center for Theoretical Physics, Randall Physics
Laboratory\\
University of Michigan, Ann Arbor, MI48109-1120, USA\\
 $^3$ Department of Physics, Osaka University, Toyonaka, Osaka
560-0043, Japan}
\end{center}
\vs{5}

\begin{abstract}
We study noncommutative open string (NCOS) theories realized in
string theory with time-dependent backgrounds. Starting from a
noncommutative Yang-Mills theory (NCYM) with a constant
space-space noncommutativity but in a time-dependent background
and making an S-dual transformation, we show that the resulting
theory is an NCOS also in a time-dependent background but now with
a time-dependent time-space noncommutativity and a time-dependent
string scale. The corresponding dual gravity description is also
given. A general $SL(2,\mathbb{Z})$ transformation on the NCYM
results in an NCOS with a time-dependent time-space
noncommutativity and a constant space-space noncommutativity, and
also in a time-dependent background.

\end{abstract}

\end{titlepage}
\newpage
\renewcommand{\thefootnote}{\arabic{footnote}}
\setcounter{footnote}{0}
\setcounter{page}{2}

%%========================section 1 ==========================
\section{Introduction}

String theory in various backgrounds is a subject of much interest.
In particular, noncommutative theories emerge from the open strings on the
D-branes in constant $B$-field background~\cite{SW}. This suggests that such
theories may be directly relevant to understanding the space-time
structure at short distances in quantum gravity. It has been known that
these theories have a very useful description in terms of gravity dual
solutions~\cite{MR,HI}, which have clarified many interesting properties
of these theories.

Most of the investigations to date are focused on static
backgrounds. It is then natural to study string theories in
time-dependent backgrounds. These theories are expected to be
important in understanding the evolution of our universe. In fact,
there have already appeared several papers on string theories in
time-dependent backgrounds~\cite{FS}-\cite{Aharony}. Various
quotients of Minkowski space-time have been studied recently as
concrete realizations of string theories on time-dependent
backgrounds. Among others, one of the simplest examples of the
space-time orbifold is that of the ``null-brane''~\cite{FS}.

The dynamics of D-branes can be studied by the attached open
strings. In a certain decoupling limit, the massive modes of open
strings as well as bulk closed strings are decoupled from the
theory, and one can study the dynamics without the complication of
gravity. It is also possible to understand the theory in terms of
the dual description by a bulk theory of gravity, in the spirit of
AdS/CFT correspondence. In particular, it has been shown that
D-branes in the null-brane background have an interesting
decoupling limit and that the resulting theory corresponds to a
noncommutative Yang-Mills theory (NCYM) with a constant
space-space noncommutativity but in a time-dependent
background\footnote{As shown in \cite{HS}, such a constant
space-space noncommutativity can be transformed back to the
original time-independent non-singular coordinate system and it
becomes dependent on both space and time.} ~\cite{HS}. The dual
description has been also given which is time-dependent. The
theory is nonlocal in space in an interesting time-dependent
manner, but no such a theory with a time-dependent
noncommutativity in the time direction and in a time-dependent
background  has been considered so far. One is then naturally led
to ask whether or not such a theory exists.

For this purpose, we study the S-dual of the above NCYM in this
paper. We show that such a theory indeed exists in our simple
setting of null-brane orbifold by using the dual gravity
description and making S-duality transformation of the solution.
We find that there is an interesting decoupling limit also in this
time-dependent case. The resulting theory turns out to be a
noncommutative open string theory (NCOS), in much the same way as
time-independent case~\cite{GMMS,SST,TH}, but now with a
time-dependent noncommutativity in the time direction and also in
a time-dependent background. We also examine the properties of the
theory under a more general $SL(2,\mathbb{Z})$ transformation. We
find that the NCOS theory is transformed again into another NCOS
in general, but under certain circumstances it is transformed into
the aforementioned NCYM, again similar to static
case~\cite{RS,JXLU,CO,LRS}.
% We start in the next section with a
%review of D3-branes in the null-brane backgrounds and the
%resulting NCYM with a constant space-space noncommutativity but in
%a time-dependent background on the worldvolume of the
%D3-branes~\cite{HS}. In Section~3, we discuss the S-duality of the
%NCYM and show that in the same decoupling limit the resulting
%theory is an NCOS also in a time-dependent background but now with
%a time-dependent time-space noncommutativity, and that a general
%$SL(2,\mathbb{Z})$ transformation gives an NCOS with both a
%time-dependent time-space noncommutativity and a constant
%space-space noncommutativity and again in a time-dependent
%background. In Section~4 we give the supergravity dual
%descriptions of other dimensional NCOS theories. We end in
%Section~5 with a brief discussion.

%%========================section2=============================
\section{D3-branes in time-dependent backgrounds and NCYM}

Let us first review the D3-branes in time-dependent null-brane
backgrounds~\cite{FS}. The geometry of the null-brane is simply an orbifold
of a Minkowski space-time
\begin{equation}
\label{2eq1}
 ds^2 =-2dx^+dx^-+dx^2 +dz^2 +dx_{\perp}^2
\end{equation}
by the identification
\begin{eqnarray}
\label{2eq2}
&& x^+ \!\! \sim x^+, \quad
x \sim x +2\pi x^+, \nn
&& x^- \!\! \sim x^-+2\pi x +(2\pi)^2 x^+/2, \quad
z \sim z +2\pi R,
\end{eqnarray}
where $dx_{\perp}^2$ denotes the line element of a six-dimensional Euclidean
space. The resulting space can also be described by the metric
\begin{equation}
\label{2eq3}
ds^2=-2d\hat x^+d\hat x^- +d\hat x^2 +(\hat x^2 +R^2)d\hat z^2
    +2(\hat x^+d\hat x -\hat x d\hat x^+)d\hat z +dx_{\perp}^2,
\end{equation}
where these coordinates have the relations to those in
(\ref{2eq1}) as follows:
\begin{eqnarray}
\hat x^+=x^+, \ \ \ \hat x^-=x^--\frac{zx}{R}+\frac{z^2x^+}{2R^2},
\ \ \ \hat x =x-\frac{zx^+}{R}, \ \ \ \hat z =\frac{z}{R}.
\end{eqnarray}
In these coordinates the orbifold becomes simple
\begin{equation}
\hat z \sim \hat z +2\pi.
\end{equation}
There is another set of coordinates~\cite{LMS}:
\begin{equation}
\label{2eq6}
x^+=y^+, \ \ \ x=y^+y, \ \ \ x^-=y^-+y^+y^2/2,
\end{equation}
in which the quotient identification is simple:
\begin{eqnarray}
&& y^+ \!\! \sim y^+, \quad
y \sim y +2\pi, \nn
&& y^- \!\! \sim y^-, \quad
z \sim z +2\pi R,
\end{eqnarray}
and the metric is also much simpler than the one (\ref{2eq3})
\begin{equation}
\label{2eq8}
 ds^2=-2dy^+dy^- +(y^+)^2dy^2 +dz^2 +dx_{\perp}^2.
 \end{equation}
Note that the coordinate transformation (\ref{2eq6}) is singular
when $y^+=0$. Since  we will adopt the coordinates in
(\ref{2eq8}), so in what follows $y^+ \ne 0$ is assumed. In the
coordinates (\ref{2eq6}) we write down the supergravity solution
of D3-branes\footnote{For the 4-form gauge potential $A_4$, we
have used a gauge transformation which removes the pure gauge term
from the 4-form potential in the usual D3-brane configuration.},
\begin{eqnarray}
\label{2eq9}
&& ds^2= H^{-1/2}(-2dy^+dy^- +(y^+)^2 dy^2 +dz^2)
    +H^{1/2} (dr^2 +r^2 d\Omega_5^2), \nonumber \\
&& A_4=\frac{y^+}{H g_s}dy^+\wedge dy^-\wedge dy\wedge dz, \qquad
e^{2\phi}=g_s^2,
\end{eqnarray}
where $g_s$ is the coupling constant of closed string and
\begin{equation}
H=1+\frac{4\pi g_s N \alpha'^2}{r^4}
\end{equation}
with $N$ being the number of D3-branes in the configuration.
Following the steps enumerated in \cite{HS}, we obtain the
required D3-brane configuration
\begin{eqnarray}
\label{2eq11}
&&  ds^2 = H^{-1/2}\left[ -2dy^+dy^- +\frac{HR^2}{HR^2 +(y^+)^2}
  \left( (y^+)^2 dy^2 +dz^2\right) \right ] +H^{1/2}(dr^2
      +r^2 d\Omega_5^2), \nonumber \\
&& 2\pi \alpha' B_{yz}=\frac{R(y^+)^2}{HR^2 +(y^+)^2}, \qquad
A_{y^+y^-}=\frac{H^{-1}}{g_s R}y^+, \nonumber \\
&& A_{y^+y^-yz}=\frac{H^{-1} y^+}{g_s}\frac{HR^2}{HR^2
       +(y^+)^2}, \qquad
e^{2\phi}=g_s^2 \frac{HR^2}{HR^2 +(y^+)^2}.
\end{eqnarray}
Taking the decoupling limit~\cite{HS}:
\begin{equation}
\label{2eq12}
\alpha' \to 0, \ \ \
u=\frac{r}{\alpha'}={\rm fixed} , \ \ \ \tilde R
=\frac{\alpha'}{R}={\rm fixed},
\end{equation}
and keeping $g_s$ constant, the supergravity configuration reduces to
\begin{eqnarray}
\label{2eq13} && ds^2 =\alpha'
\frac{u^2}{\lambda^2}\left[-2dy^+dy^-
  +h^{-1}((y^+)^2 dy^2 +dz^2) +\frac{\lambda^4}{u^4}(du^2
  +u^2d\Omega_5^2)\right], \nonumber \\
&& 2\pi \alpha' B_{yz}=\alpha' \frac{\tilde R (y^+)^2 u^4}{\lambda^4}
h^{-1}, \qquad A_{y^+y^-}=\alpha' \frac{\tilde  R y^+ u^4}{g_s
\lambda^4}, \qquad e^{2\phi}= g_s^2 h^{-1},
\end{eqnarray}
where $\lambda ^4= 4 \pi g_s  N$, and
$$ h =1 +\frac{\tilde R^2 (y^+)^2 u^4}{\lambda ^4}.$$
It is easy to confirm that under the decoupling limit
(\ref{2eq12}), the resulting theory is indeed a noncommutative
Yang-Mills theory in ($3+1)$ dimensions. Naively using the
Seiberg-Witten relation~\cite{SW}
%\begin{eqnarray}
%\label{2eq14}
%&& G_{ij}=g_{ij}-(2\pi \alpha')(Bg^{-1}B)_{ij}, \nonumber \\
%&& \Theta ^{ij}=2\pi \alpha' \left( \frac{1}{g +2\pi \alpha'
%   B }\right)^{ij}_A, \nonumber \\
%&& G^{ij}=\left (\frac{1}{g+2\pi\alpha' B}\right)^{ij}_S, \nonumber \\
%&& G_s=e^{\phi}\left(\frac{\det G_{ij}}{\det(g_{ij}+2\pi \alpha'
%   B_{ij})}\right)^{1/2},
%\end{eqnarray}
to the solution (\ref{2eq11}) in the ``flat'' limit $H=1$, we get
the open string moduli $G_s =g_s$,
\begin{equation}
\label{ssnop} \Theta^{yz}=  2\pi \tilde R,
\end{equation}
and
\begin{equation}
\label{2eq15}
G^{ij}= \left (
\begin{array}{cccc}
0 & -1& 0& 0 \\
-1& 0& 0 &  0 \\
0 &0& (y^+)^{-2} & 0\\
0 & 0 & 0 & 1
\end{array} \right).
\end{equation}
This indicates that we have a good open string moduli with {\it
constant} noncommutativity parameter in the coordinates
(\ref{2eq15}), although the closed string metrics and hence the
theory itself is time-dependent~\cite{HS}. It is quite interesting
that even though the closed string metrics is time-dependent, they
are so in an intricate way that the noncommutativity parameter is
constant. However, it is worthwhile  mentioning here that as shown
in \cite{HS}, if one uses the coordinates (\ref{2eq1}), the
Yang-Mills theory lives in a flat static space with noncommutative
parameter depending on space-time. We will see that the situation
is drastically changed for our time-dependent NCOS theories.

%%======================section 3===============================
\section{S-duality and NCOS}

In \cite{GMMS} it is shown that strongly coupled, spatially
noncommutative ${\cal N}=4$ Yang-Mills theory has a dual
description as a weakly coupled noncommutative open string theory.
What is the case of D3-branes in time-dependent backgrounds?

Under the S-duality,
%\footnote{To be consistent with what we will
%discuss later in this section on $SL(2, \mathbb{Z})$-duality, we choose the
%S-duality to leave the string constant $\alpha'$ and the Einstein
%metric invariant. The original string metric is asymptotically flat.}
the supergravity dual (\ref{2eq13}) becomes\footnote{In obtaining
the following, we actually rescale the coordinates such that the
various fields have the desired dependences on the $g_s$ factor
which is convenient for us to discuss the corresponding NCOS in
other dimensions. They are: $y^+ \to g_s^{1/2} y^+,\, y^- \to
g_s^{-3/2} y^-,\, y\to g_s^{-1} y,\, z\to g_s^{- 1/2} z,\, u \to
g_s^{- 1/2} u,\, R\to g_s^{1/2} R, \, \tilde R \to g_s^{-
1/2}\tilde R$. This is not important in the present case but it
will be for other D$p$-branes discussed in section 4 since the
$g_s$ factor scales in the corresponding decoupling limit.}
\begin{eqnarray}
\label{3eq1}
&& ds^2 =\alpha' \frac{u^2}{\lambda^2} h^{1/2} \left[-2dy^+dy^-
  +h^{-1}((y^+)^2 dy^2 +dz^2) +\frac{\lambda ^4}{u^4}(du^2
  +u^2d\Omega_5^2)\right], \nonumber \\
&&2\pi \alpha' B_{y^+y^-}=
  \alpha' \frac{ \tilde  R y^+ u^4}{\lambda 4}, \qquad
A_{yz}=-\alpha' \frac{\tilde R (y^+)^2 u^4}{g_s \lambda ^4}h^{-1},
\qquad e^{2\phi}= g_s^2 h,
\end{eqnarray}
where $\lambda^4 = 4\pi g_s N$ again.

What is the dual (open string) theory to the supergravity
configuration~(\ref{3eq1})? To see this, let us first write
down\footnote{We also rescale the coordinates as in footnote 3
with the replacement of $u \to g_s^{- 1/2} u$ by $r \to g_s^{-
1/2} r$.} the S-dual of the solution (\ref{2eq11}):
\begin{eqnarray}
\label{3eq2}
&& ds^2 =  H^{-1/2}F^{-1/2}\left[-2dy^+dy^-+F((y^+)^2
   dy^2 +dz^2) +H (dr^2 +r^2 d\Omega_5^2)\right], \nonumber \\
&& 2\pi \a' B_{y^+y^-}=\frac{ y^+}{HR}, \qquad A_{yz}=
-\frac{(y^+)^2 F}{g_s RH}, \qquad e^{2\phi}=g_s^2 F^{-1},
\end{eqnarray}
where $$ F=\frac{HR^2}{HR^2 +(y^+)^2},$$ and $g_s$ is the inverse
of the original string coupling. Taking the decoupling limit
(\ref{2eq12}),  the resulting configuration  from the solution
(\ref{3eq2}) is identical to that of (\ref{3eq1}).

In the ``flat'' space limit $H=1$, we can examine the open string
moduli by using Seiberg-Witten relation again. From eq.~\p{3eq2} with $H=1$,
we find the open string parameters are given by
\bea
G^{ij} =
\left(
\begin{array}{cccc}
0 & -\sqrt{\frac{R^2+(y^+)^2}{R^2}} & 0& 0 \\
-\sqrt{\frac{R^2+(y^+)^2}{R^2}} & 0& 0 &  0 \\
0 & 0 & \sqrt{\frac{R^2+(y^+)^2}{R^2}}\frac{1}{(y^+)^2} & 0\\
0 & 0 & 0 & \sqrt{\frac{R^2+(y^+)^2}{R^2}}
\end{array} \right),
\ena
and
 \bea \Theta^{y^+ y^-}= 2 \pi \a'\frac{y^+}{R} = 2 \pi y^+ \tilde R.
\label{noncp} \ena Note that the noncommutativity
parameter~\p{noncp} is well-defined in the decoupling
limit~\p{2eq12} but it is now time-dependent, in contrast to the
space-space noncommutative case~\p{ssnop} for the NCYM! Also $\a'
G^{ij}$ is finite in the decoupling limit~\p{2eq12}, which means
that although the modes of closed strings decouple in the limit
(\ref{2eq12}), the massive modes from the open strings do not
decouple, resulting in a noncommutative open string theory.
Further, it is easy to find that in that case the effective open
string scale is $\alpha'_{\rm eff}= y^+ \tilde R$, depending on
time and the coupling constant of open strings is still a
constant, $G_s =g_s$.

The supergravity description \p{3eq1} may be compared with the one
dual to NCOS  with constant time-space noncommutativity given in
\cite{GMMS}. Write \bea h &=& \frac{u^4}{A^4} \left( 1+
\frac{A^4}{u^4} \right)\nn &\equiv& \frac{u^4}{A^4} f(u),\quad A^4
\equiv \frac{\lambda ^4}{\tilde R^2 (y^+)^2}, \ena and we get \bea
&& ds^2 = \alpha'  f^{1/2} \left[\frac{u^4}{\lambda ^2
A^2}(-2dy^+dy^-)
 + \frac{A^2}{\lambda^2} f^{-1}((y^+)^2 dy^2 +dz^2)
 +\frac{\lambda^2}{A^2} (du^2 +u^2d\Omega_5^2)\right], \nn
&& 2\pi \alpha' B_{y^+y^-}=
 \alpha' \frac{ \tilde  R y^+ u^4}{\lambda ^4}, \qquad
A_{yz}=-\alpha' \frac{1}{ g_s \tilde R f}, \qquad e^{2\phi}= g_s^2
\frac{u^4}{A^4}f. \ena If $A^4$ were constant, this would be
exactly the gravity solution dual to NCOS with constant
noncommutativity~\cite{GMMS}. For small $u$, we recover the AdS$_5
\times$S$^5$ but with an orbifolding of the flat 4-dimensional
slice in $AdS_5$ since the open string theory reduces to $N=2$
super Yang-Mills theory at low energies or long distance ($N=2$ due to
orbifolding). However, the theory significantly deviates from that
for $u \sim A$, which determines the size of the noncommutativity.

The solution has actually time-dependent noncommutativity, and
along constant light-cone coordinate $y^+$, the theory looks
exactly as the NCOS with constant noncommutativity. Away from that
region, the theory has different noncommutativity scale.

More generally we can consider an $SL(2,\mathbb{Z})$
transformation to the solution (\ref{2eq11}). Since the axion
field is zero, we have \bea \tilde\tau = \frac{a \tau +b}{c\tau
+d}, \quad \tau \equiv i e^{-\phi},\quad ad-bc =1,\quad a,b,c,d
\in \mathbb{Z}, \ena whose imaginary part gives \bea
e^{-\tilde\phi} = \frac{e^{-\phi}}{|c\tau + d|^2}. \ena The
Einstein-frame metric is unchanged under the $SL(2,\mathbb{Z})$,
so $ds_E^2 = e^{-\phi/2} ds^2= e^{-\tilde\phi/2} d{\tilde s}^2$,
giving \bea d{\tilde s}^2= |c\tau+d| ds^2. \ena We find from the
solution~\p{2eq11} the transformed configuration as\footnote{In
obtaining the following configuration, we have as before rescaled
the coordinates as: $y^+ \to g_s^{-1/2}(c^2 + d^2 g_s^2)^{1/4}
y^+,\; y^- \to g_s^{3/2}(c^2 + d^2 g_s^2)^{-3/4} y^-, \; y\to
g_s(c^2 + d^2g_s^2)^{-1/2} y, \; z\to g_s^{1/2}(c^2 + d^2
g_s^2)^{-1/4} z, \; r\to g_s^{1/2}(c^2 + d^2 g_s^2)^{-1/4} r, \;
R\to g_s^{- 1/2}(c^2 + d^2 g_s^2)^{1/4} R$. Note that the string
coupling $g_s$ here is the original one while the one used in
footnotes 3 and 4 is the transformed one.  }
 \bea && d{\tilde s}^2 =
F^{-1/2} H^{-1/2}\sqrt{\frac{c^2 + g_s^2 d^2 F}{c^2 + g_s^2 d^2} }
\Bigg[ -2 dy^+ dy^- + F\,
 ((y^+)^2 d y^2 + dz^2)  \nn
  && \hs{12} +H(dr^2+r^2 d\Omega_5^2) \Bigg], \nn
&& 2\pi \alpha' \tilde B = \frac{d(y^+)^2 F}{H R}
\frac{g_s}{\sqrt{c^2 + d^2 g_s^2}} dy\wedge dz - \frac{c y^+}{ HR}
\frac{1}{\sqrt{c^2 + d^2 g_s^2}} dy^+ \wedge dy^-, \nn &&
e^{2\tilde\phi} = g_s^{-2} F
   \left ( c^2 +d^2 g_s^2 F\right)^2, \,\,
 \tilde\chi =\left (ac +bd g_s^2 F\right)
   \left( c^2 +d^2 g_s^2 F \right)^{-1},
 \label{sl2ts}
\ena where the function $F$ is the same as before, $g_s$ is the
original string coupling and ${\tilde g}_s = g_s^{-1} (c^2 + d^2
g_s^2)$ is the new one. The harmonic function $H = 1 + 4\pi
\alpha'^2 \tilde g_s N/r^4$.

Using the Seiberg-Witten relation, we obtain the open string
moduli: the open string metric
\begin{equation}
G^{ij}= \left (
\begin{array}{cccc}
0 &-\sqrt{\frac{g_s^2 d^2 +c^2 w}{c^2 + g_s^2 d^2}} &0&0 \\
-\sqrt{\frac{g_s^2 d^2 +c^2 w}{c^2 + g_s^2 d^2}} &0&0&0 \\
0&0& \frac{1}{(y^+)^2}\sqrt{\frac{g_s^2 d^2 +c^2 w}{c^2 + g_s^2 d^2}}&0 \\
0&0&0&\sqrt{\frac{g_s^2 d^2 +c^2 w}{c^2 + g_s^2 d^2}}
\end{array} \right ),
\end{equation}
the noncommutative parameter
\begin{equation}
\Theta^{ij}= 2\pi \alpha' \left (
\begin{array}{cccc}
0 & - \frac{cy^+}{R \sqrt{c^2 + g_s^2 d^2}} &0 &0\\
\frac{cy^+}{R \sqrt{c^2 + g_s^2 d^2}} &0&0&0\\
0&0&0& - \frac{g_s d}{R \sqrt{c^2 + g_s^2 d^2} }\\
0&0&\frac{g_s d}{R \sqrt{c^2 + g_s^2 d^2} }&0
\end{array} \right ),
\end{equation}
and the open string coupling constant $G_s = {\tilde g}_s =
g_s^{-1} (c^2 + g_s^2 d^2)$. In the above, the function $w = 1 +
(y^+)^2/R^2$. One can check that the above moduli give the correct
ones when $a, b, c, d$ are specialized to their corresponding values.

In the decoupling limit (\ref{2eq12}), assuming $c\ne 0$, we get
\bea
&& \hs{-10}
d{\tilde s}^2 = \alpha' \sqrt{\frac{\tilde f}{1 + g_s^2
d^2/c^2}} \left[\frac{u^4}{\lambda^2 A^2} (-2dy^+dy^-)
 + \frac{A^2}{\lambda ^2} f^{-1}((y^+)^2 dy^2 +dz^2)
 +\frac{\lambda^2}{A^2} (du^2 +u^2d\Omega_5^2)\right],\nn
&& 2\pi\a' {\tilde B} = \a'\frac{1}{f\tilde R}\frac{d
g_s}{\sqrt{c^2 + d^2 g_s^2}} dy \wedge dz
 - \a'\frac{\tilde R u^4 y^+}{\lambda^4}\frac{c}{\sqrt{c^2 + d^2 g_s^2}}
  dy^+ \wedge dy^-, \nn
&& e^{2\tilde \phi} =g_s^{- 2} \left (1+\frac{u^4}{A^4}\right) \left( c^2 +
  \frac{g_s^2 d^2}{1+u^4/A^4}\right)^2, \nn
&& \tilde\chi =\left ( ac +\frac{g_s^2 bd}{1+u^4/A^4}\right) \left (c^2
 +\frac{g_s^2 d^2}{1+u^4/A^4}\right)^{-1},
\ena
where
\bea
\tilde f = 1+ \frac{c^2+ g_s^2 d^2}{c^2 u^4} A^4.
\ena
This corresponds to a general NCOS theory with a
time-dependent  time-space noncommutativity and a constant
space-space noncommutativity. The effective string scale is now
$\alpha'_{\rm eff}= y^+ \tilde R c/\sqrt{c^2 + d^2 g_s^2}$. The
space-space noncommutativity is directly proportional to the
original string coupling $g_s$ if $c \ne 0$. For the special case
of $c=0$, it gives a gravity dual to NCYM. So we can see
from~\p{sl2ts} that in the case of $c=0$, the space-space
noncommutative NCYM in a time-dependent background is transformed
back to NCYM. Thus we find that the space-space noncommutative
NCYM theory in a time-dependent background is transformed into
NCOS in general, but for the special case $c=0$ it is transformed
to NCYM again. Since we get an NCOS theory from the NCYM by an
$SL(2,\mathbb{Z})$ transformation, which makes a group, by making
a further  $SL(2,\mathbb{Z})$ transformation to the obtained NCOS
theory, we get another NCOS theory. This means that the NCOS
theory itself transforms into NCOS in general, but of course it may
transform into NCYM in special case. This is much the same as the
time-independent noncommutative theories~\cite{CO,LRS}.

%%=========================section 4===============================

\section{The Cases for Other dimensions}

The authors of \cite{AP} have extended the study in \cite{HS} to
other dimensional D$p$-branes, NS$5$-branes and M$5$-branes in
time-dependent backgrounds, and discussed the supergravity duals
of decoupled worldvolume theories. In the previous section we have
discussed supergravity dual of ($3+1$)-dimensional NCYM with
space-space noncommutativity but in a time-dependent background
and its S-duality, resulting in NCOS with time-dependent
noncommutativity. In this section we give the supergravity dual of
other dimensional NCOS theory.

Applying T-duality to the solution (\ref{3eq2}), we can obtain
D$p$-brane solution with electric field in the ``null-brane"
geometry which can be viewed as deformed (F, D$p$) bound states
discussed in \cite{LR},
\begin{eqnarray}
\label{4eq1} && ds^2= H^{-1/2}F^{-1/2}[-2dy^+dy^-+F((y^+)^2dy^2
+dz^2 +\sum^{p-3}_{i=1}dx_i^2) +H(dr^2 +r^2d\Omega_{8-p}^2)],
     \nonumber\\
&& 2\pi \alpha' B_{y^+y^-}=\frac{ y^+}{HR}, \ ~~~~~~ \ \ e^{2\phi}
     =g_s^2H^{\frac{3-p}{2}}F^{\frac{p-5}{2}},
     \nonumber \\
&& A_{yzx_1\cdots x_{p-3}}=-\frac{(y^+)^2 F}{g_s RH},
  \ ~~~~~~~ \ \ A_{y^+y^-yzx_1\cdots x_{p-3}}=
    \frac{ y^+F}{H g_s},
\end{eqnarray}
where
$$ F=\frac{HR^2}{HR^2 +(y^+)^2}, \ \ \
H=1+\frac{c_pg_sN\alpha'^{(7-p)/2}}{r^{7-p}} $$
with $c_p=2^{5-p}\pi^{\frac{5-p}{2}}\Gamma[(7-p)/2]$. Considering
the following decoupling limit
\begin{equation}
\label{4eq2} \alpha' \to 0, \ \ \ \bar g_s = g_s
{\alpha'}^{(p-3)/2}={\rm fixed},\ \ \ u =\frac{r}{\alpha'}={\rm
fixed}, \ \ \ \tilde R=\frac{\alpha'}{R}={\rm fixed},
\end{equation}
the supergravity solution becomes
\begin{eqnarray}
\label{4eq3}
&& ds^2=\alpha' \left(\frac{u}{\lambda}\right)^{(7-p)/2}h^{1/2}
    \left (-2dy^+dy^- +h^{-1} ((y^+)^2dy^2 +dz^2
    +\sum^{p-3}_{i=1}dx_i^2) \right. \nonumber \\
&&~~~~~~\left. +\left(\frac{\lambda}{u}\right)^{7-p}
    (du^2 +u^2 d\Omega_{8-p}^2)\right),  \nonumber \\
&& 2\pi \alpha' B_{y^+y^-}=\alpha' y^+\tilde R \left(\frac{u}{\lambda}
   \right)^{7-p}, \ \ \
e^{2\phi}= \bar
   g_s^2h^{(5-p)/2}\left(\frac{\lambda}{u}\right)^{(7-p)(3-p)/2},
\end{eqnarray}
where $$ h =1 +\frac{\tilde R^2 (y^+)^2 u^{7-p}}{\lambda ^{7-p}}$$
and $\lambda^{7-p}= c_p \bar g_s N$. Naively using Seiberg-Witten
relation, we can easily show that the worldvolume theory on the
D$p$-brane ($p<6$) in the solution (\ref{4eq1}) decouples from the
closed string theory in the decoupling limit (\ref{4eq2}),
resulting in a $(p+1)$-dimensioanl NCOS theory with time-dependent
noncommutativity and open string scale $\alpha'_{\rm eff}= \tilde
R y^+$. The open string coupling constant $G_s = g_s (R^2/(R^2 +
(y^+)^2))^{(p - 3)/4} = \bar g_s (\tilde R y^+)^{(3-p)/2}$.
So only for $p = 3$, the $G_s$ is time-independent. The supergravity
solution (\ref{4eq3}) is just the gravity dual description of the
decoupled NCOS theory within the region where the supergravity
description is valid in the usual manner~\cite{Mald}.

%%=============================secttion 5==========================
\section{Conclusions}

In this paper we have discussed time-dependent noncommutative
theories obtained from string theories in time-dependent
background on the simple null-brane orbifold. The resulting NCYM
theory has only a constant space-space noncommutativity but in a
time-dependent background.  Using S-duality transformation, we
have identified the resulting theory as a noncommutative NCOS in a
time-dependent background but with a time-dependent time-space
noncommutativity. We have also examined the transformation
properties of the NCYM under a general $SL(2,\mathbb{Z})$
transformation, and find that it is transformed into NCOS in
general, but under a special circumstance it is transformed back
to NCYM . This is quite similar to the case of noncommutative
theories resulting from D-branes in the static backgrounds.

%{\it \bf It would be quite interesting to describe NCOS theories
%in the coordinates (\ref{2eq1}).}

\section*{Acknowledgement}
%\noindent{\large\bf Acknowledgments}:
The work of R.G.C. was supported in part by a grant from Chinese
Academy of Sciences. J.X.L. acknowledges also the partial support of
a grant from the Chinese Academy of Sciences. The work of N.O. was
supported in part by a Grant-in-Aid for Scientific Research No. 12640270.

\end{document}